\documentstyle[eqsecnum,aps]{revtex}
\begin{document}
\draft
\title{CHARACTERISTICS OF COSMIC TIME}
\author{D. S. Salopek}
\vspace{2pc}
\address{
{}Department of Physics, University of Alberta,
Edmonton, Canada T6G 2J1 }
\vspace{2pc}

\date{May 26, 1995}

\maketitle

\begin{abstract}
The nature of cosmic time is illuminated using
Hamilton-Jacobi theory for general relativity.
For problems of interest to cosmology, one may solve
for the phase of the wavefunctional by using a line integral
in superspace. Each contour of integration corresponds to a
particular choice of time hypersurface, and each yields
the same answer. In this way, one can construct
a covariant formalism where all time hypersurfaces are treated
on an equal footing. Using the method of characteristics,
explicit solutions for an inflationary epoch with
several scalar fields are given. The theoretical predictions
of double inflation are compared with recent galaxy data and large
angle microwave background anisotropies.

\end{abstract}

\pacs{\hfill ALBERTA THY/09-95}


\widetext

\section{Introduction}

One obtains a better appreciation for the nature of cosmic time by
studying  the role of inhomogeneities in general relativity
\cite{SS95}-\cite{SB}. After all, a time-hypersurface represents
the arbitrary manner in which one slices a 4-geometry.
However, there are numerous problems associated with quantum aspects
of time \cite{KUCHAR}. One should perhaps be content to consider the
semiclassical limit of the quantum theory
where one retains  only the leading order terms in Planck's constant
$\hbar$ that appear in the Wheeler-DeWitt equation.
This approximation leads directly to the
Hamilton-Jacobi (HJ) equation for general relativity. In the present work,
superspace solution techniques of the HJ equation will be
further advanced to include the case of several scalar fields
interacting in a cosmological setting. These methods simplify
the comparison of theoretical models with astronomical observations.

The problem of choosing a gauge in general
relativity has complicated the interpretation
as well as the application of the theory.
For example, when solving Einstein's equations,
one must make arbitrary choices for the
space-time coordinates.  Many of these decisions may be postponed
or even avoided by solving the HJ equation for
general relativity.  Hamilton-Jacobi theory leads to a covariant
description of the gravitational field \cite{SS95}-\cite{SB}.
It yields a generalization of earlier work on covariant cosmological
perturbations by Ellis {\it et al} \cite{EHB89}.

Already, HJ techniques have been applied to
numerous problems in cosmology including: (1) a detailed computation
of microwave background fluctuations and galaxy-galaxy
correlations \cite{SS95}, \cite{S92}  generated in the power-law inflation
model \cite{LM85} which  arises either in induced gravity \cite{SBB89} or
extended inflation \cite{LS90};
(2) a relativistic approach to the Zel'dovich approximation
\cite{Zel} describing the formation of sheet-like structures
during the matter-dominated era \cite{CPSS94}, \cite{SSC94},
\cite{MPS93};
(3) attempts to recover the inflaton potential from
cosmological observations \cite{SBB89}, \cite{COPELAND93};
(4) a construction of inflation models that yield
non-Gaussian primordial fluctuations \cite{S91} ---- such models
may alleviate the problems of large scale structure in the Universe
\cite{Mosc93}.

In Sec. II, I set forth the HJ equation and the momentum
constraint equation describing the interaction of two scalar
fields with the gravitational field. The extension to more
fields is straightforward. The object of chief
importance is the generating functional
${\cal S}$ which is the phase of the wavefunctional in the
semiclassical approximation.  In Sec. III, a
spatial gradient expansion is used to solve the HJ equation.
The zeroth order term for the generating
functional, ${\cal S}^{(0)}$, contains
no spatial gradients. It describes the evolution of long-wavelength
fields. The second term, ${\cal S}^{(2)}$, contains two spatial
derivatives. Its evolution equation is a linear partial
differential equation which may be simplified using the method
of characteristics. In particular, by invoking a local change of variables,
one may solve this equation by using a line integral in superspace.
Different contours of integration correspond to different
choices of the time-hypersurface, but they all lead to the
same answer for ${\cal S}^{(2)}$. In Sec. IV, it is shown how to
effectively sum an infinite subset of terms in the spatial gradient
expansion. This  generalizes the quadratic curvature approximation that was
employed by Salopek and Stewart \cite{SS95}. The resulting Riccati
equations may be reduced to ordinary linear differential equations using
matrix methods. In Sec. V, initial conditions
are given for a period of cosmological inflation. In Sec. VI,
some examples and solutions are given for inflation with two scalar
fields. Using analytic methods, one is able to compute fluctuation
spectra that give rise to galaxy clustering. Previously, one
required a tedious numerical computation \cite{SBB89}.
Theoretical models are constrained by large angle microwave
background anisotropies and galaxy clustering data.

(Units are chosen so that $c=8\pi G=8\pi / m_P^2= \hbar= 1$.
The  sign conventions of Misner, Thorne and
Wheeler \cite{MTW} will be adopted throughout.)

\section{The Hamilton-Jacobi equation for general relativity}

For two scalar fields $\phi_a$, $a=1,2$, interacting through
Einstein gravity, the Hamilton-Jacobi equation,
\begin{mathletters}
\begin{eqnarray}
0={\cal H}(x)&&=\gamma^{-1/2}
\left[2\gamma_{ik}(x) \gamma_{jl}(x) - \gamma_{ij}(x)\gamma_{kl}(x)\right]
{\delta{\cal S}\over \delta\gamma_{ij}(x)}
{\delta{\cal S}\over \delta\gamma_{kl}(x)} + \nonumber\\
&& \sum_{a=1}^{2}
{1\over 2} \gamma^{-1/2}\left({\delta{\cal S}\over \delta\phi_a(x)}\right)^2
+\gamma^{1/2} V[\phi_a(x)] + \;
\left [ -{1\over 2}\gamma^{1/2}R
+ \sum_{a=1}^{2}
{1\over 2} \gamma^{1/2}\gamma^{ij}\phi_{a,i}\phi_{a,j} \right ] \,  ,
\label{HJES}
\end{eqnarray}
and the momentum constraint equation,
\begin{equation}
0={\cal H}_{i}(x)=
-2\left(\gamma_{ik}{\delta{\cal S}\over \delta\gamma_{kj}(x)}
\right)_{,j} +
{\delta{\cal S}\over\delta\gamma_{kl}(x)}\gamma_{kl,i} +
\sum_{a=1}^{2} {\delta{\cal S}\over\delta\phi_a (x)} \phi_{a,i}  \, ,
\label{MCS}
\end{equation}
\end{mathletters}
govern the evolution of the generating functional,
${\cal S}\equiv {\cal S}[\phi_a(x), \gamma_{ij}(x)]$, in
{\it superspace}.
For each field configuration $\phi_a(x)$ on a space-like hypersurface with
3-geometry described by the 3-metric $\gamma_{ij}(x)$, the generating
functional associates a complex number.
Here $R$ denotes the Ricci scalar of the 3-metric, and
$V[\phi_a(x)]$ is the scalar field potential.
In the ADM formalism, the space-time line element is written as
\begin{equation}
ds^2\equiv g_{\mu \nu} dx^\mu dx^\nu =
\left(-N^2+\gamma^{ij}N_iN_j\right)dt^2 + 2N_i dt\,dx^i +
\gamma_{ij}dx^i\,dx^j\ ,
\label{ADMdecomp}
\end{equation}
where $N$ and $N_i$ are the lapse and shift functions, respectively.
The generating functional is
the `phase' of the wavefunctional in the semiclassical approximation,
\begin{mathletters}
\begin{equation}
\Psi \sim e^{i{\cal S}} \, . \label{phase}
\end{equation}
(The prefactor is neglected here although it has important implications
for quantum cosmology \cite{B93}.)
The probability functional
\begin{equation}
{\cal P} \equiv |\Psi|^2,  \label{prob}
\end{equation}
\end{mathletters}
is just the square of the wavefunctional (see, e.g., ref. \cite{S92b}).

The HJ equation (\ref{HJES}) and the momentum
constraint (\ref{MCS}) follow, respectively, from the
$G^0_0$ and $G^0_i$ Einstein equations with the canonical
momenta replaced by functional derivatives of ${\cal S}$:
\begin{equation}
\pi^{\phi_a}(x) = { \delta {\cal S} \over \delta \phi_a(x) } \, ,
\quad {\rm and} \quad
\pi^{ij}(x) = { \delta {\cal S} \over \delta \gamma_{ij}(x) } \, .
\end{equation}
The momentum constraint demands that the generating functional be
invariant under an arbitrary change of the spatial coordinates \cite{Peres}
(see also ref.\cite{MTW}, p.1185). The Hamilton-Jacobi equation for general
relativity is analogous to the
Tomonaga-Schwinger equation that yielded a covariant
formulation of quantum electrodynamics \cite{Tomo.Schw}.
If the generating functional is real,  the evolution of the 3-metric
for one particular universe is given by
\begin{mathletters}
\begin{equation}
\left(\dot\gamma_{ij}-N_{i|j}-N_{j|i}\right)/N =2\gamma^{-1/2}
\left(2\gamma_{jk}\gamma_{il}-\gamma_{ij}\gamma_{kl}\right)
{ \delta {\cal S} \over \delta \gamma_{kl} } \, ,
\label{evol.metric}
\end{equation}
whereas the evolution equation for the scalar field is
\begin{equation}
\left(\dot\phi_a-N^i\phi_{a,i}\right)/N=\gamma^{-1/2}
{\delta {\cal S} \over \delta \phi_a} \, .
\label{evol.scalar}
\end{equation}
\end{mathletters}
Here $|$ denotes a covariant derivative with respect to the 3-metric
$\gamma_{ij}$. Since the lapse and shift function appear neither
in the HJ equation (\ref{HJES}) nor in the momentum constraint (\ref{MCS}),
these two equations are {\it covariant} in that they are valid for all
choices of the space-time coordinates.
All gauge-dependent quantities which are associated with the lapse and shift
appear only in the evolution equations (\ref{evol.metric},b)
for the 3-metric and scalar field.

\section{SPATIAL GRADIENT EXPANSION OF THE GENERATING FUNCTIONAL}

Superspace describes an ensemble of evolving universes, and its
complexity is overwhelming. One may reduce this complexity by
expanding the generating functional
\begin{equation}
{\cal S}= \sum_{n=0}^{\infty} {\cal S}^{(2n)}
\label{theexpansion}
\end{equation}
in a series of terms according to
the number of spatial gradients that they contain.
The great simplification that arises is that
one encounters linear partial differential  equations at
each order for
$n \ge 1$. A careful exposition is given of the generating
functional of order two. This example illustrates many of the
essential aspects required in solving more challenging problems.
The consistency of higher order solutions is also demonstrated.

\subsection{Generating Functional of Order Zero}

To order zero, the Hamilton-Jacobi equation is
\begin{eqnarray}
\gamma^{-1/2} \left(2\gamma_{ik}\gamma_{jl}- \gamma_{ij}\gamma_{kl}\right)
{\delta {\cal S}\over \delta\gamma_{ij}}^{(0)}
{\delta{\cal S}\over \delta\gamma_{kl}}^{(0)}
+ \sum_{a=1}^2 {1\over 2} \gamma^{-1/2}\left(
{\delta{\cal S}\over \delta\phi_a}^{(0)}\right)^2
 + \gamma^{1/2}V\left(\phi_a\right) =0.
\label{zeroham}
\end{eqnarray}
In the full HJ eq.(\ref{HJES}), the spatial derivative terms
appearing in the square brackets  have been neglected.
They will be recovered at the next order. One attempts a
solution of the form,
\begin{mathletters}
\begin{equation}
{\cal S}^{(0)}[\phi_a(x), \gamma_{ij}(x)]
=-2\int d^3x\,\gamma^{1/2}\, H \left ( \phi_a(x) \right ).
\label{zerothsol}
\end{equation}
for the zeroth order generating functional.
The numerical factor $-2$ is chosen so that the
function of the scalar fields, $H\equiv H(\phi_a)$, corresponds to
the usual Hubble parameter in the long-wavelength approximation.
Because the integral is over $d^3x\,\gamma^{1/2}$,  this
functional is invariant under spatial coordinate transformations, and
consequently the momentum constraint eq.(\ref{MCS}) is satisfied.
The HJ eq.(\ref{zeroham}) of order zero holds provided that $H$ satisfies
the following nonlinear partial differential equation:
\begin{equation}
H^2={2\over 3}\sum_{a=1}^2 \left({\partial H\over\partial\phi_a}\right)^2
   +{1 \over 3}V\left(\phi_a\right)\ .
\label{SHJE}
\end{equation}
\end{mathletters}
Since the metric  does not appear in this equation,
it is often referred to as the separated Hamilton-Jacobi equation
(SHJE) (of order zero).

\subsection{Generating Functional of Order Two}

The second order HJ equation is
\begin{equation}
\sum_{a=1}^2
\left [ -2{\partial{H}\over\partial\phi_a}{\delta{\cal
S}^{(2)}\over\delta\phi_a}
\right ]
+ 2H\gamma_{ij} {\delta{\cal S}^{(2)}\over\delta\gamma_{ij}}
= {1\over 2}\gamma^{1/2}R
    - \sum_{a=1}^2 {1\over 2}\gamma^{1/2}\gamma^{ij}\phi_{a,i}\phi_{a,j}  \,
{}.
\label{secondham}
\end{equation}
It is a linear partial differential equation of the
inhomogeneous type. It may be solved using a two step
method. (1) One firstly applies the method of characteristics
to choose an integration parameter or time parameter;
the remainder of the variables are then frozen in time. (2) Secondly,
following the work of Salopek and Stewart \cite{SS92}, one makes
an ansatz for the second order generating functional
which contains all terms with two spatial derivatives.
If one wishes to be more direct, one may employ
a line integral in superspace \cite{PSS94} which elegantly illuminates
the nature of cosmic time.

The characteristic equations
\begin{mathletters}
\begin{eqnarray}
{ 1 \over N} { \partial \phi_a \over \partial t} =&&
             -2 { \partial H \over \partial \phi_a}  \, , \\
{ 1 \over N} { \partial \gamma_{ij} \over \partial t} =&& 2 H \gamma_{ij}
\end{eqnarray}
\end{mathletters}
are determined by the coefficients
of the functional derivatives, $\delta {\cal S}/ \delta \phi_a$
and $\delta {\cal S}/ \delta \gamma_{ij}$,
appearing in eq.(\ref{secondham}).
Here the time parameter $t$ as well as the lapse $N$ are arbitrary,
but for the sake of simplicity one can safely assume that
they are local functions of the scalar fields: $t\equiv t(\phi_a)$,
$N \equiv N(\phi_a)$.
The above equations may be simplified by performing
a conformal transformation,
\begin{equation}
\gamma_{ij}(x) = \Omega^2(\phi_a) \,  f_{ij}(x) \, ,
\end{equation}
where the conformal factor, $\Omega \equiv \Omega (\phi_a)$, is
a function of the scalar fields, and the conformal 3-metric,
$f_{ij}$, is independent of time $t$.
The reduced characteristic equations are then:
\begin{mathletters}
\begin{eqnarray}
{ 1 \over N} { \partial \phi_a \over \partial t} =&&
             -2 { \partial H \over \partial \phi_a}  \, ,
\label{chred1} \\
{ 1 \over N} { \partial \ln \Omega \over \partial t} =&&  H  \, .
\label{chred2}
\end{eqnarray}
\end{mathletters}

There are numerous ways to solve these equations.
If one is fortunate to obtain a solution for the
Hubble function $H \equiv H(\phi_a, \widetilde \phi_a)$ which
depends on two homogeneous and time-independent
parameters $\widetilde \phi_a$, $a=1,2$, then
one may integrate these equations immediately:
\begin{mathletters}
\begin{eqnarray}
\Omega(\phi_a) =&& \left( { \partial H \over \partial \widetilde \phi_1}
                   \right )^{-1/3} \, ,  \label{txa} \\
{e}(\phi_a) =&& { \partial H \over \partial \widetilde \phi_1}/
            { \partial H \over \partial \widetilde \phi_2} \,  , \label{txb} \\
f_{ij} =&& \Omega^{-2}(\phi_a) \;  \gamma_{ij} \, . \label{txc}
\end{eqnarray}
\end{mathletters}
Both ${e}(\phi_a)$ and $f_{ij}$ are independent of time:
\begin{mathletters}
\begin{eqnarray}
{ 1\over N} { \partial e \over \partial t} &&= 0 \, , \\
{ 1\over N} { \partial f_{ij} \over \partial t} &&= 0 \, .
\end{eqnarray}
\end{mathletters}
 By differentiating the SHJE (\ref{SHJE}) with respect to the parameters,
one may verify that these fields satisfy the reduced
characteristic equations (\ref{chred2}).  Note that this solution
refers neither to the time parameter $t$ nor to the lapse $N$---
they are redundant variables.

The integration of the characteristic equations may be viewed
as a transformation of the fields
\begin{equation}
(\phi_1, \phi_2, \gamma_{ij} ) \rightarrow (\Omega, {e},  f_{ij} ).
\end{equation}
Utilizing the conformal 3-metric $f_{ij}$ instead
the original 3-metric $\gamma_{ij}$ is analogous
to using comoving coordinates rather than physical
coordinates in cosmological systems.
$\Omega(x) \equiv \Omega[\phi_a(x)]$ can be interpreted as the
integration parameter or `time'
parameter, whereas ${e(x)}, f_{ij}(x)$ are independent of time.
The integration parameter could have been chosen to be any function of the
scalar fields, but the choice of $t= \Omega \equiv \Omega(\phi_a)$ will
simplify the subsequent integration. From the second
characteristic equation (\ref{chred2}),
$\partial \Omega / \partial t=1$, implies that the lapse is given by
$N = 1/(\Omega H)$, and hence the scalar field evolves
in $\Omega$ according to
\begin{equation}
\left( { \partial \phi_a \over \partial \Omega} \right)_{e} =
- { 2 \over \Omega H} \, { \partial H \over \partial \phi_a} \, .
\label{che}
\end{equation}
In some instances, it may be difficult to determine a solution
for the Hubble function which depends on two parameters.
Alternatively, one may integrate eq.(\ref{che}) where
${e}$ now refers to an arbitrary and inhomogeneous constant of
integration.

Using the new variables, the HJ equation of order two may be rewritten as
\begin{equation}
\Omega H \, {\delta{\cal S}^{(2)}\over\delta\Omega} \Bigg
|_{{e}, f_{ij}} = {1\over 2 } \gamma^{1/2} R
- \sum_{a=1}^2 {1\over 2 }\gamma^{1/2}\gamma^{ij}\phi_{a,i}\phi_{a,j} \, .
\label{secondham2}
\end{equation}
The method of characteristics has simplified the left hand side
considerably. It now remains to express the right hand side in
terms of the new variables. For example,
a conformal transformation of the Ricci curvature leads to
\begin{equation}
R(\Omega^2 f_{ij}) = \Omega^{-2} \widetilde R
-8 \Omega^{-5/2} (\Omega^{1/2})^{;i}_{;i}
\end{equation}
where $\widetilde R$ is the Ricci scalar of the conformal 3-metric
$f_{ij}$, and a semi-colon $;$ denotes a covariant derivative
with respect to the conformal 3-metric. As a result, one finds
the following final form for the HJ equation of order two,
\begin{mathletters}
\begin{equation}
{ \delta{\cal S}^{(2)} \over\delta\Omega } \Bigg |_{{e}, f_{ij}}
= f^{1/2} \left [
   { \widetilde R \over 2 H }
- { 2 \over H \Omega } \Omega^{;i}_{;i} +
    l(\Omega, {e}) \,  \Omega_{;i} \Omega^{;i}
+ 2 m(\Omega, {e}) \,  \Omega_{;i} {e}^{;i}
+   n(\Omega, {e}) {e}_{;i} {e}^{;i}
          \right ] \label{fd1}
\end{equation}
where $l, m, n$ are functions of $(\Omega, {e})$:
\begin{eqnarray}
l(\Omega, {e}) =&&  { 1 \over H \Omega^2} - { 1 \over 2 H}
\left [ \left (  { \partial \phi_1 \over \partial \Omega} \right )_{e}^2 +
        \left (  { \partial \phi_2 \over \partial \Omega} \right )_{e}^2
\right ]  \, , \\
m(\Omega, {e}) =&&  -{ 1 \over 2 H}
\left[  \left ( { \partial \phi_1 \over \partial \Omega} \right)_{e}
        \left ( {\partial \phi_1 \over \partial {e}}\right)_\Omega +
        \left ( { \partial \phi_2 \over \partial \Omega} \right)_{e}
        \left ( { \partial \phi_2 \over \partial {e}}  \right)_\Omega
\right ] \, , \\
n(\Omega, {e}) =&&  - {1 \over 2 H}
\left [ \left (  { \partial \phi_1 \over \partial {e}} \right )_\Omega^2 +
        \left (  { \partial \phi_2 \over \partial {e}} \right )_\Omega^2
\right ] \, .
\label{eqe}
\end{eqnarray}
\end{mathletters}
\subsubsection{Ansatz for Second Order Solution}

An Ansatz,
\begin{mathletters}
\begin{equation}
{\cal S}^{(2)} = \int d^3x \, f^{1/2} \, \left [
j(\Omega, {e}) \widetilde R +
k_{11}(\Omega, {e}) \Omega_{;i} \Omega^{;i} +
2k_{12}(\Omega, {e}) \Omega_{;i} {e}^{;i} +
k_{22}(\Omega, {e}) {e}_{;i} {e}^{;i} \right ] \, .
\label{ansatzp}
\end{equation}
will be used to integrate the HJ equation of order two.
$j$, $k_{11}$, $k_{12}$ and $k_{22}$,  are functions of
$(\Omega, {e})$ which are determined to be:
\begin{eqnarray}
j(\Omega, {e})=&&
\int_0^\Omega d \Omega^\prime \; { 1 \over 2 H(\Omega^\prime, {e})} + j_0(e) \,
,\\
k_{11}(\Omega, {e})=&& { 1 \over H \Omega} \, ,\\
k_{22}(\Omega, {e})=&&\int_0^\Omega d \Omega^\prime \;  n(\Omega^\prime, {e})
+ k_0(e) \, , \\
k_{12}(\Omega, {e})=&& 0 \, ,
\end{eqnarray}
\end{mathletters}
where $n(\Omega, {e})$ was defined in eq.(\ref{eqe}), and
$j_0 \equiv j_0(e) $ and $k_0 \equiv k_0(e)$ are arbitrary functions
of $e$.

In order to construct the above solution, one computes the functional
derivative
with respect to $\Omega(x)$ of the Ansatz eq.(\ref{ansatzp}):
\begin{equation}
f^{-1/2} \, { \delta{\cal S}^{(2)} \over\delta\Omega } \Bigg |_{{e}, f_{ij}}
= {\partial j \over \partial \Omega}  \widetilde R
- {\partial k_{11} \over \partial \Omega}  \Omega_{;i} \Omega^{;i}
- 2 {\partial k_{11} \over \partial {e}}  \Omega_{;i} {e}^{;i}
+ \left( { \partial k_{22} \over \partial \Omega}
- 2 {\partial k_{12} \over \partial {e}}
\right) {e}_{;i} {e}^{;i} -2 k_{11} \Omega^{;i}_{;i}
-2 k_{12} {e}^{;i}_{;i} \, ,
\end{equation}
and matches terms with HJ equation (\ref{fd1}) of order two.
Since no ${e}^{;i}_{;i}$ term appears in eq.(\ref{fd1}), one finds
immediately that the cross term $k_{12}$ vanishes. This fortunate
circumstance arose because we chose $\Omega$ to be our
integration parameter. The remaining equations
\begin{mathletters}
\begin{eqnarray}
\left ( {\partial j \over \partial \Omega} \right )_e
=&& { 1 \over 2 H} \, , \label{evj} \\
k_{11}=&& { 1 \over H \Omega } \, , \label{k11d} \\
\left ( {\partial k_{11} \over \partial \Omega} \right )_e =&& -l \, ,
\label{k11e1} \\
\left ( {\partial k_{11} \over \partial {e}} \right )_\Omega =&&-m \,
,\label{k11e2}\\
\left ( {\partial k_{22} \over \partial \Omega} \right )_e =&&  n \, ,
\label{k22e}
\end{eqnarray}
\end{mathletters}
lead to the solution given above.
Initially, it appears that this system of equations may be
overdetermined since  $k_{11}$ appears in the three equations
(\ref{k11d}-\ref{k11e2}). It is a minor miracle
that given the first equation, the latter two are automatically
satisfied.

The integration of the HJ equation of order two is a nontrivial
result. The ansatz (\ref{ansatzp}) and the prescription
for determining its free functions represent
a complete and explicit integration of the problem.
Previously, Salopek and Stewart \cite{SS92}
had solved the single field system.
The new ingredients required for resolving
the two scalar field problem were:
(1) the method of characteristics, and
(2) choosing $\Omega(x) \equiv \Omega[\phi_a(x)]$
as the integration parameter. Even then, the final
form, eqs.(\ref{ansatzp}-e), of the generating functional of order two is
valid for {\it all} choices of the time hypersurface.

\subsubsection{Line Integrals and Potential Theory }

The HJ eq.(\ref{fd1}) of order two has the form of an infinite
dimensional gradient. It may be integrated directly using a line
integral in superspace. The required aspects of potential theory are
first briefly reviewed.

The fundamental problem in potential theory is: given a force
field $g^i(y_k)$ which is a function of $n$ variables $y_k$,
what is the potential $\Phi \equiv \Phi(y_k)$ (if it exists)
whose gradient returns the force field:
\begin{equation}
{\partial \Phi \over \partial y_i} = g^i(y_k) \quad ?
\end{equation}
Not all force fields are derivable from
a potential. Provided that the force field satisfies the
integrability relation,
\begin{equation}
0= {\partial g^i \over \partial y_j} - {\partial g^j \over \partial y_i} =
\left [{\partial  \over \partial y_j}, {\partial  \over  \partial y_i }
\right ] \, \Phi \, , \label{curl}
\end{equation}
(i.e., it is curl-free),
one may find a solution which is conveniently expressed using a
line integral
\begin{equation}
\Phi(y_k) = \int_C \sum_j d {\overline y}_j \ g^j( {\overline y}_l) \ .
\end{equation}
If the two endpoints are fixed, all contours return the same
answer. In practice, we will employ the simplest contour that
one can imagine: a line connecting the origin to the
observation point $y_k$. Using $s$, $0 \le s \le 1$,
to parameterize the contour, ${\overline y}_l = s y_l$,
the line integral may be rewritten as
\begin{equation}
\Phi(y_k) = \sum_{j=1}^n \int_0^1 ds  \; y_j \ g^j( {\overline y}_l) +
\Phi_0
\ ,
\end{equation}
where $\Phi_0$ is independent of $y_k$.

\subsubsection{Line Integral Method for Second Order Generating Functional}

By identifying, $j \rightarrow x$, $\sum_j \rightarrow \int d^3x$,
$y_j \rightarrow \Omega(x)$, $dy_j = ds \Omega(x) $,
one may integrate the infinite dimensional gradient eq.(\ref{fd1}),
\begin{equation}
{\cal S}^{(2)}[\Omega(x), {e}(x), f_{ij}(x)]= \int d^3x \int_0^1 ds \;
\Omega(x) {\delta{\cal S}^{(2)} \over\delta {\overline \Omega} } \Bigg |_{{e},
f_{ij}}
+{\cal S}^{(2)}_0[{e}(x), f_{ij}(x)] \, ,
\label{lint}
\end{equation}
with ${\overline \Omega}(x) = s \Omega(x)$. One recovers
the result (\ref{ansatzp}) guessed earlier
provided the `constant functional', ${\cal S}^{(2)}_0$, is given by
\begin{equation}
{\cal S}^{(2)}_0[{e}(x), f_{ij}(x)] =  \int d^3 x f^{1/2} \,
\left [ j_0(e) \widetilde R + k_0(e) f^{ij} e_{,i} e_{,j} \right ]  \, ,
\end{equation}
which is independent of $\Omega(x)$.
Because of an integrability condition discussed in the
next section, this result does not depend on the contour of integration
in superspace.

\subsubsection{Integrability Condition for Generating Functional}

A line integral proves useful for computing higher
order terms in the spatial gradient expansion.

For $n \ge 1$, one finds that $S^{(2n)}$ satisfies the following
inhomogeneous, linear partial differential equation:
\begin{mathletters}
\begin{equation}
\widehat O (x) \; {\cal S}^{(2n)} = -{\cal R}^{(2n)}(x) \, .
\label{hoa}
\end{equation}
The differential operator $\widehat O(x)$ is given by
\begin{equation}
\widehat O(x) \equiv
\sum_{a=1}^2 -2 {\partial{H}\over \partial\phi_a}
{\delta \over \delta \phi_a(x)} +
2H \gamma_{ij}
{\delta \over\delta\gamma_{ij}(x)}
\, . \label{hob}
\end{equation}
The remainder term ${\cal R}^{(2n)}(x)$,
\begin{eqnarray}
 {\cal R}^{(2n)}(x) =&&
  \gamma^{-1/2}\sum_{p=1}^{n-1}
 {\delta{\cal S}^{(2p)}\over\delta\gamma_{ij}(x) }
 {\delta{\cal S}^{(2n-2p)}\over\delta\gamma_{kl}(x) }
 \left(2\gamma_{jk}\gamma_{li}- \gamma_{ij}\gamma_{kl}\right) \nonumber \\
+&& \gamma^{-1/2}\sum_{p=1}^{n-1}
\sum_{a=1}^2  {1\over 2} {\delta{\cal S}^{(2p)} \over\delta\phi_a(x)}
  {\delta{\cal S}^{(2n-2p)}\over\delta\phi_a(x)} + {\cal V}^{(2n)} \, .
\end{eqnarray}
is independent of ${\cal S}^{(2n)}$ but it contains contributions
from all previous orders --- it is assumed to be known.
The superspace potential ${\cal V}^{(2n)}$ is defined to be
\begin{eqnarray}
{\cal V}^{(2n)}(x) =&&
-{1\over 2}\gamma^{1/2} R  +
\sum_{a=1}^2 {1\over 2}\gamma^{1/2}\gamma^{ij}\phi_{a,i}\phi_{a,j} \quad
\rm {for \; n=1\; ,} \nonumber \\
=&& 0 \quad \rm{otherwise.}
\end{eqnarray}
\end{mathletters}
By manipulating eq.(\ref{hoa}-d), one may compute the following commutator,
\begin{eqnarray}
\left [ \widehat O(x), \widehat O(y) \right ] {\cal S}^{(2n)} \equiv &&
\widehat O(y)  \;  {\cal R}^{(2n)}(x) - \widehat O(x) \;  {\cal R}^{(2n)}(y)
\nonumber \\
=&& [ \gamma^{ij}(x) \; {\cal H}^{(2n-2)}_j(x)+
\gamma^{ij}(x^\prime) \; {\cal H}^{(2n-2)}_j(x^\prime) ] \;
{ \partial \over \partial x^i } \delta^3  (x - x^\prime) \, ,
\label{comm}
\end{eqnarray}
which assumes by induction that ${\cal S}^{(2)}, {\cal S}^{(4)}, \ldots,
{\cal S}^{(2n-2)}$ satisfy eq.({\ref{hoa}).
The `integrability condition' of potential theory, eq.(\ref{curl}),
demands that the commutator (\ref{comm}) vanish.
In the above expression, $ {\cal H}^{(2n-2)}_j $ is the momentum constraint
evaluated using the generating functional of order $(2n-2)$:
\begin{equation}
{\cal H}^{(2n-2)}_j(x)\equiv
-2\left(\gamma_{jk}{\delta{\cal S}^{(2n-2)} \over \delta\gamma_{kl}(x)}
\right)_{,l} +
{\delta{\cal S}^{(2n-2)}\over\delta\gamma_{kl}(x)}\gamma_{kl,j} +
\sum_{a=1}^{2} {\delta{\cal S}^{(2n-2)}\over\delta\phi_a (x)} \phi_{a,i}  \, .
\end{equation}
We conclude that ${\cal S}^{(2n)}$ is indeed integrable provided
the term of previous order,
${\cal S}^{(2n-2)}$, is invariant under reparametrizations of the
spatial coordinates: ${\cal H}^{(2n-2)}_j=0$.
In general, the integrability condition for the Hamilton-Jacobi equation
follows from  the Poisson brackets \cite{MT72} between the energy densities
evaluated at the two spatial points $x$ and $x^\prime$:
\begin{equation}
\{ {\cal H}(x), {\cal H}(x^\prime) \} =
[ \gamma^{ij}(x) {\cal H}_j(x)+
\gamma^{ij}(x^\prime) {\cal H}_j(x^\prime) ] \;
{ \partial \over \partial x^i } \delta^3  (x - x^\prime) \, .
\label{poisson}
\end{equation}

In practice, one would solve the HJ eq.(\ref{hoa}) of order 2n
by changing variables according to eq.(\ref{txa}-c), and then
applying a line integral analogous to eq.(\ref{lint}):
\begin{eqnarray}
{\cal S}^{(2n)}[\Omega(x), {e}(x), f_{ij}(x)]=&& -\int d^3x  \int_0^1
{ ds \over s }  \;
{ 1 \over H \left [ {\overline \Omega}(x), e(x)  \right ]  }  \;
{\cal R}^{(2n)}[ {\overline \Omega}(x), {e}(x), f_{ij}(x)]   \\
&& +{\cal S}^{(2n)}_0[{e}(x), f_{ij}(x)] \, , \quad \quad{\rm with} \;
{\overline \Omega}(x) = s\,  \Omega(x) \, .
\label{lintn}
\end{eqnarray}
Once again, the `constant functional,' ${\cal S}^{(2n)}_0[{e}(x), f_{ij}(x)]$
depends only on the fields $e(x)$, $f_{ij}(x)$, which
are independent of $\Omega(x)$.
The constant functional contains all such terms with $2n$ spatial
derivatives which are invariant under reparametrizations of the
spatial coordinates.

\subsection{The Nature of Cosmic Time}

For the semiclassical limit, a line integral in
superspace goes a long way towards illuminating the nature of cosmic time.
Different contours of integration analogous to eq.(\ref{lintn})
correspond to different intermediate choices for the integration
parameter or time parameter. However, they all
yield the same answer for the generating functional provided
that spatial gauge-invariance is maintained at each order of
the computation.

\section{QUADRATIC CONSTANT APPROXIMATION FOR
GRAVITY WITH MULTIPLE SCALAR FIELDS}

In the last section, a prescription was given for computing
the generating functional to arbitrary order in the spatial
gradient expansion. For a single scalar field or a single
dust field, this has been done explicitly to 4th order
or 6th order, respectively, in
spatial gradients \cite{PSS94}. The case for a single dust field
is of considerable practical importance because it leads
to the Zel'dovich approximation for general relativity
\cite{CPSS94}, \cite{SSC94}.

However, for some cosmological applications, a finite number of
terms is insufficient. In this section, it will be shown
how to effectively sum an infinite subset of terms in the
spatial gradient expansion. The `quadratic constant
approximation' generalizes the `quadratic curvature
approximation' developed in an earlier paper
\cite{SS95}.

\subsection{Factoring Out the Long-Wavelength Background}

Beginning with the full HJ eq.({\ref{HJES}}), one first subtracts
the long-wavelength background from the generating functional:
\begin{equation}
{\cal S} = {\cal S}^{(0)} + {\cal F} \ , \quad
{\cal S}^{(0)} = - 2 \int d^3x \gamma^{1/2} H(\phi_a) \ ,
\label{subtract}
\end{equation}
where $H$ satisfies the separated Hamilton-Jacobi equation (\ref{SHJE}).
The functional for fluctuations, $\cal F$, now satisfies
\begin{eqnarray}
&&
\sum_{a=1}^2 -2 { \partial H \over \partial \phi_a}
{\delta {\cal F} \over \delta \phi_a} +
2 H \gamma_{ij} { \delta {\cal F} \over \delta \gamma_{ij} } +
\gamma^{-1/2} \left[2\gamma_{il}(x) \gamma_{jk}(x) -
\gamma_{ij}(x)\gamma_{kl}(x)\right]
{\delta{\cal F}\over \delta \gamma_{ij}(x)}
{\delta{\cal F}\over \delta\gamma_{kl}(x)} \nonumber \\
&& + {1\over 2} \gamma^{-1/2}
\sum_{a=1}^2 \left({\delta{\cal F}\over \delta\phi_a(x)}\right)^2
-{1\over 2}\gamma^{1/2}R
+\sum_{a=1}^2 {1\over 2} \gamma^{1/2}\gamma^{ij}\phi_{a,i}\phi_{a,j}=0\ .
\label{HJE1}
\end{eqnarray}
The first line of eq.(\ref{HJE1}) may simplified if one introduces
the change of variables, $(\phi_a, \gamma_{ij}) \rightarrow
(\Omega, {e}, f_{ij} )$, described by eqs.(\ref{txa}-c).
Functional derivatives with respect to the fields transform according to
\begin{equation}
{ \delta \over \delta \gamma_{ij} } = \Omega^{-2}  \;
{ \delta \over \delta f_{ij} } \Bigg |_{\Omega, {e}} \ , \quad
{\delta \over \delta \phi_a} =
{ \partial \Omega \over \partial \phi_a}
\left ( \;{ \delta \over \delta \Omega}\Bigg |_{{e}, f_{ij}}
- {2 \over \Omega} f_{ij} { \delta \over \delta f_{ij} }\Bigg |_{\Omega, {e}}
\; \right  )  +
{\partial {e} \over \partial \phi_a}  { \delta \over \delta {e}}
\Bigg|_{\Omega, f_{ij}} \,  \ .
\end{equation}
(Henceforth, one
suppresses the symbols $|_\Omega$, $|_{e}$, $|_{f_{ij}}$ which
denote the variables that are held constant during
functional differentiation.)

The HJ equation reduces to
\begin{eqnarray}
{ \delta {\cal F} \over \delta \Omega }+
{ f^{-1/2} \over \Omega^4 H} \left[2 f_{il} f_{jk} - f_{ij} f_{kl}\right]
{\delta{\cal F}\over \delta f_{ij}} {\delta{\cal F}\over \delta f_{kl}}
+ \sum_{a=1}^2 { f^{-1/2} \over 2\Omega^{4} H} \,
\left[
{ \partial \Omega \over \partial \phi_a}
\left ( \;{ \delta {\cal F} \over \delta \Omega}
- {2 \over \Omega} f_{ij} { \delta {\cal F} \over \delta f_{ij} }
\; \right  )  +
{\partial {e} \over \partial \phi_a}  { \delta {\cal F} \over \delta {e}}
\right]^2
=  { \delta {\cal S}^{(2)} \over \delta \Omega }
\label{HJEP}
\end{eqnarray}
where ${\cal S}^{(2)}$ was given in eq.(\ref{ansatzp}).
The momentum constraint maintains the same form as before
but it is now expressed in terms of the new variables
$(\Omega, {e}, f_{ij})$:
\begin{equation}
{\cal H}_{i}(x)=-2\left(f_{ik}{\delta{\cal F}\over \delta f_{kj}}
\right)_{,j} +
{\delta{\cal F}\over\delta f_{kl}} f_{kl,i}
+ {\delta{\cal F}\over \delta {e} } {e}_{,i}
+ {\delta{\cal F}\over \delta \Omega } \Omega_{,i}=0 \ \ .
\label{Snewmomentum}
\end{equation}

\subsection{Integral form of HJ equation}

Following ref.\cite{PSS94}, an  integral form of the
HJ eq.(\ref{HJEP}) may be constructed using a line integral in superspace,
\begin{eqnarray}
{\cal F}&&[ \Omega(x), {e}(x), f_{ij}(x) ] +
\int d^3x \int_0^1 { ds \over s }  { f^{-1/2} \over  {\overline \Omega}^{3} H }
\left[2 f_{il} f_{jk} - f_{ij} f_{kl} \right]
{ \delta{\cal F} \over \delta f_{ij} }
{ \delta{\cal F} \over \delta f_{kl} }
+ \nonumber \\
&& \int d^3 x \int_0^1 { ds \over s }
\sum_{a=1}^2 {f^{-1/2} \over {\overline \Omega}^{3} H}
\left[ { \partial \Omega \over \partial \phi_a}
\left(  { \delta {\cal F} \over \delta {\overline \Omega} }
- 2 {\overline \Omega}^{-1} f_{ij}
{ \delta{\cal F} \over \delta f_{ij} } \right ) +
{ \partial {e} \over \partial \phi_a} { \delta {\cal F} \over \delta {e}}
\right]^2
=  {\cal S}^{(2)}[\Omega(x), {e}(x), f_{ij}(x) ] \, ,
\label{IHJE}
\end{eqnarray}
where once again ${\overline \Omega}(x) = s \Omega(x)$.

\subsection{Ansatz}

One makes an Ansatz for the fluctuation functional (\ref{subtract}) of the
form,
\begin{equation}
{\cal F} = {\cal S}^{(2)}  + {\cal Q} \, ,
\label{ansatz1}
\end{equation}
where the {\it quadratic functional},
\begin{equation}
{ \cal Q} = \int d^3x f^{1/2} \left [
\widetilde R \, \widehat S_{11} \, \widetilde R +
2 \widetilde R \, \widehat S_{12} \, (\widetilde D^2 {e})+
(\widetilde D^2 {e}) \, \widehat S_{22} \, (\widetilde D^2 {e}) +
 \widetilde R^{ij} \, \widehat T \, \widetilde R_{ij}
- { 3 \over 8 }   \widetilde R \, \widehat T \, \widetilde R
\right ] \, ,
\label{ansatz2}
\end{equation}
consists of all quadratic combinations of the fields,
$\widetilde D^2 e$, $\widetilde R$, $\widetilde R_{ij}$
(but {\it not} $\, \widetilde D^2 \Omega$),
which maintain spatial gauge invariance.
$\widetilde D^2$ is the Laplacian operator
with respect to the conformal 3-metric, e.g.,
\begin{equation}
\widetilde D^2 \widetilde R \equiv \widetilde D^{;i} \widetilde D_{;i} \,
\widetilde R
\equiv   \widetilde R^{;i}{}_{;i}
= f^{-1/2} \left( f^{1/2} f^{ij} \widetilde R_{,j} \right )_{,i} \ ,
\end{equation}
One interprets the operator $\widehat T$ for tensor
perturbations to be a Taylor series of the form
\begin{equation}
\widehat T(\Omega, {e}, \widetilde D^2) = \sum_n T_n(\Omega, {e})
(\widetilde D^2)^n  \, ,
\end{equation}
and similarly for the scalar operator $\widehat S_{ab}$,
which is a two-by-two symmetric matrix.
(The full Riemann tensor $\widetilde R_{ijkl}$ does not appear in the Ansatz
because for three spatial dimensions because it may be written in terms of
the Ricci tensor \cite{SSP93}.)
Although the first and fifth terms in eq.(\ref{ansatz2}) may be
combined into a single one, the present form simplifies
the final evolution equations for $\widehat S$ and $\widehat T$
(see eqs.(\ref{ev1}), (\ref{ev2})). The Ansatz (\ref{ansatz2}) generalizes
the quadratic curvature approximation considered by
Salopek and Stewart \cite{SS95}.

The perturbation rules are the same as those
developed in ref.\cite{SS95}.
By first order, we refer to terms such as $\widetilde R$,
$\widetilde D^2 \Omega$, $\widetilde D^2 {e}$,
or  $ \widetilde D^2 \widetilde  R$, $\widetilde D^4 \Omega$,
$\widetilde D^4 {e}$, which vanish if the fields
are homogeneous; they may contain any number of
spatial derivatives. Quadratic terms are a product of
two linear terms.

In computing the various functional
derivatives, it is useful to note that for
a small variation of the conformal 3-metric $\delta f_{ij}$
the corresponding change in the Ricci tensor is
\begin{equation}
\delta \widetilde R_{ij} = {1 \over 2 } f^{kl} \left [
\delta f_{li;jk} + \delta f_{lj;ik}- \delta f_{ij;lk}
- \delta f_{lk;ij} \right ] \, .
\label{small}
\end{equation}
In the integral form of the HJ equation, integration by parts
is permitted which simplies the analysis considerably.
(However, in its differential form (\ref{HJES})
one cannot simply discard total spatial derivatives.)
In addition, all cubic terms are neglected.

To linear order, the functional derivatives of ${\cal S}^{(2)}$,
eq.(\ref{ansatzp}), are
\begin{mathletters}
\begin{eqnarray}
f^{-1/2} \, {\delta { \cal S}^{(2)} \over \delta \Omega} =&&
{\partial j \over \partial \Omega} \, \widetilde R - 2 k_{11} \,
\widetilde D^2 \Omega \, \\
f^{-1/2} \, {\delta { \cal S}^{(2)} \over \delta f_{ij}} =&&
j \, \left ( {\widetilde R \over 2} f^{ij} - \widetilde R^{ij} \right )
+ j^{;ij} - j^{;k}_{;k} f^{ij} \, , \\
f^{-1/2} \, {\delta { \cal S}^{(2)} \over \delta {e}} =&&
{\partial j \over \partial {e}} \, \widetilde R - 2 k_{22} \,
\widetilde D^2 {e} \, .
\end{eqnarray}
\end{mathletters}
and those of the quadratic functional, eq.(\ref{ansatz2}), are
\begin{mathletters}
\begin{eqnarray}
f^{-1/2} \, {\delta {\cal Q} \over \delta \Omega}   =&& 0  \, , \\
f^{-1/2} \, {\delta { \cal Q} \over \delta f_{ij} } =&&
2 \widehat S_{11} \left [ \widetilde R^{;ij}
- \widetilde D^2 \widetilde R f^{ij} \right ] +
2 \widehat S_{12}  \left [ \left( \widetilde D^2 {e} \right )^{;ij} -
\widetilde D^4 {e}  f^{ij} \right ]  +     \\
&&\widehat T \left [ { 1\over 4 } \widetilde R^{;ij} +
{ 1\over 4 } \widetilde D^2 \widetilde R f^{ij}-
\widetilde D^2 \widetilde R^{ij}  \right ] \, , \\
f^{-1/2} \, {\delta { \cal Q} \over \delta {e}}       =&&
2 \widetilde D^2 \left [ \widehat S_{12} \widetilde R +
\widehat S_{22} \widetilde D^2 {e} \right ]  \, .
\end{eqnarray}
\end{mathletters}
At the present level of approximation,  $\widetilde D^2$
commutes with any function of $(\Omega, E)$, {\it e.g.},
$\widetilde D^2 S_{11} = S_{11} \widetilde D^2$,   {\it etc}.
After a straightforward computation
(which is very similar to that of ref.\cite{SS95}), one obtains
the  tensor equation
\begin{equation}
{\partial \widehat T \over \partial \ln \Omega} +
{2 \over  \Omega^3 H }
\left ( j + \widehat T \widetilde D^2 \right )^2  =0 \, ,
\quad {\rm (tensor)}
\label{TRICCATI}
\end{equation}
and the scalar equation
\begin{mathletters}
\begin{equation}
{\partial \widehat S \over \partial \ln \Omega} +
2 \widetilde D^4 \; \widehat S \widehat A \widehat S
+ \widetilde D^2 \; \widehat C^{T} \widehat S
+ \widetilde D^2 \; \widehat S \widehat C +
{1 \over 2} \widehat B=0 \, ,
\quad {\rm (scalar)} \label{SRICCATI}
\end{equation}
where $\widehat A$, $\widehat B$ and $\widehat C$ are operator matrices.
(From now on, the operator symbol $\widehat{}$ will be suppressed.)
They are given by
\begin{eqnarray}
A=&& { 1 \over  \Omega^3 H}
\left [ \vec a_1  {\vec a_1}^{T} + \vec a_2  {\vec a_2}^{T}  \right ] \, ,
\label{mata} \\
B=&& { 1 \over  \Omega^3 H}
\left [ \vec b_1  {\vec b_1}^{T} + \vec b_2  {\vec b_2}^{T} \right ] \, ,\\
C=&& { 1 \over  \Omega^3 H}
\left [ \vec a_1  {\vec b_1}^{T} + \vec a_2   {\vec b_2}^{T}  \right ] \, .
\end{eqnarray}
Here ${T}$ denotes the transpose of a matrix or a vector, and the
vectors $\vec a_1$, $\vec a_2$ and $\vec b_1$, $\vec b_2$ are
\begin{equation}
{\vec a_1}^{T} = \left [ 4  { \partial \ln \Omega \over \partial \phi_1}, \;
                { \partial {e} \over \partial \phi_1} \right ] \, ,
\end{equation}
\begin{equation}
{\vec a_2}^{T} = \left [ 4 { \partial \ln \Omega \over \partial \phi_2}, \;
                { \partial {e} \over \partial \phi_2} \right ] \, ,
\end{equation}
\begin{equation}
{\vec b_1}^{T} = \left [ \left ( { \partial  \Omega \over \partial \phi_1}
\right )
\left ( {\partial j \over \partial \Omega}  - {j \over \Omega} \right )+
{\partial j \over \partial {e}} {\partial {e} \over \partial \phi_1} \, , \;
4 {\partial \ln \Omega \over \partial \phi_1} { \partial j \over \partial {e}}
- 2 k_{22} { \partial {e} \over \partial \phi_1} \right ] \, ,
\end{equation}
\begin{equation}
{\vec b_2}^{T} = \left [ \left ( { \partial  \Omega \over \partial \phi_2}
\right )
\left ( {\partial j \over \partial \Omega}  - {j \over \Omega} \right )+
{\partial j \over \partial {e}} {\partial {e} \over \partial \phi_2} \, , \;
4 {\partial \ln \Omega \over \partial \phi_2} { \partial j \over \partial {e}}
- 2 k_{22}  { \partial {e} \over \partial \phi_2} \right ] \, ,
\end{equation}
\end{mathletters}
The tensor equation is a single nonlinear ordinary differential equation
of the Riccati type. The scalar equation is also of the Riccati type, but
the dependent variable $S_{ab}$ is a 2-by-2 symmetric matrix.
An additional complications arise in the latter case because the
coefficients $A$, $B$, $C$, appearing in eq.(\ref{SRICCATI})
need not be commuting.

\subsection{Reduction of Riccati Equations}

The Riccati equations, (\ref{TRICCATI}) and
(\ref{SRICCATI}), may be reduced to linear ordinary differential equations.
For the tensor perturbation, one defines the Riccati transformation,
$T(\Omega, e, \widetilde D^2) \rightarrow y(\Omega, e, \widetilde D^2)$,
\begin{mathletters}
\begin{equation}
\widehat T = { H \Omega^3 \over 2 \widetilde D^4} \
{ 1 \over y} { \partial y \over \partial \ln \Omega} - \widetilde D^{-2} j \, ,
\label{ricct}  \quad  {\rm (tensor \; perturbation)}
\end{equation}
which leads to the linear equation for a massless field :
\begin{equation}
0 = { \partial ^2 y \over \partial (\ln \Omega)^2} +
\left [ 3 + { 1 \over H } { \partial H \over \partial \ln \Omega} \right ]
{\partial y \over \partial \ln \Omega}
-  {1 \over H^2 \Omega^2  } \widetilde D^2 y \, .
\quad {\rm (tensor\; perturbation)}
\label{ev1}
\end{equation}
\end{mathletters}

Because non-commuting matrices appear in the scalar case,
one must be careful about the order of various matrices, but
otherwise the method is the same as the previous case.
By defining the matrix Riccati transformation,
$ S_{ab}(\Omega, {e}, \widetilde D^2) \rightarrow
W_{ab}(\Omega, {e}, \widetilde D^2)$,
\begin{mathletters}
\begin{equation}
S = {1 \over 2 \widetilde D^4} \; A^{-1}
\; { \partial W \over \partial \ln \Omega} \;  W^{-1}
- {1 \over 2 \widetilde D^2} A^{-1} C \, ,
\quad  {\rm (scalar \; perturbation)}
\label{riccs}
\end{equation}
one recovers the linear matrix equation for the scalar perturbation,
\begin{equation}
0 = {\partial ^2 W \over \partial ( \ln \Omega)^2 }
+ A  { \partial A^{-1} \over \partial \ln \Omega} \;   \;
{\partial  W \over \partial \ln \Omega}
 -  {1 \over H^2 \Omega^2  } \widetilde D^2 W \, .
 \quad  {\rm (scalar \; perturbation)}
\label{ev2}
\end{equation}
\end{mathletters}

The case of a single scalar field has been much studied.
Mukhanov {\it et al} \cite{Mukhanov92} derived the scalar
perturbation equation by considering quadratic perturbations
of the Einstein action. Hawking, Laflamme and Lyons \cite{HLL93} utilized
this equation in their discussion of the arrow of time. Using the Einstein
field equations, Hwang \cite{Hwang93} gave an elegant interpretation
in terms of uniform curvature gauge.
Salopek and Stewart \cite{SS95} used the scalar equation to discuss
microwave anisotropies and galaxy correlations arising from
inflation.
Deruelle {\it et al} \cite{DGL92} employed
an equation which is basically equivalent to eq.(\ref{ev2}) in their discussion
of extended inflation \cite{LS90} which incorporates two scalar
fields. A path integral approach for multiple fields was
considered in ref.\cite{AM94}. However, the present form of the scalar
perturbation equation (\ref{ev2}) possesses some distinct advantages
over earlier formulations.

Because there are no mass terms appearing on the right-hand-side,
eq.({\ref{ev2}) is a particularly elegant form for the scalar
perturbation equation. For example, in the large wavelength limit,
one may neglect the
last term appearing on the right-hand-side of eq.(\ref{ev2}).
For an inflationary epoch, $W$ then approaches a constant,
\begin{equation}
W(\Omega, e, \tilde D^2) \rightarrow W^\infty(e, \tilde D^2) \,
\end{equation}
({\it i.e.}, it is independent of the integration parameter $\ln \Omega$).
The numerical value of $W^\infty$ and how it
depends on $\widetilde D^2$ are of high importance
for observational cosmology. Approximate results
will be given in the next section.

\section{INITIAL CONDITIONS}

It is difficult to obtain analytic solutions of the
scalar perturbation equation (\ref{ev2}) for two scalar fields.
Numerical computations have been considered in an earlier
work \cite{SBB89}. An approximate analytic method which
is well known for the case of a single scalar field case will be extended
to include multiple fields.  Long wavelength and short wavelength perturbations
will be matched at the Hubble radius.
(Solutions of the tensor evolution equation
(\ref{ev1}) and their applications to cosmology have already been examined,
\cite{SS95},\cite{S92}, and they will not be discussed further in this
section.)

During a period of cosmological inflation, the initial conditions
for the tensor perturbation $y$ and the scalar perturbation matrix $W$
are set by quantum conditions: for wavelengths that are much shorter
than the Hubble radius, $1/ H$, one assumes that each mode began
in the ground state.

It is useful to rewrite the scalar evolution equation
as a set of first order differential equations.
One first introduces a momentum variable $P$, which
is a matrix defined by
\begin{equation}
P = A^{-1} \; {\partial  W \over \partial \ln \Omega} \, .
\end{equation}
The evolution equation for the pair $(W, P)$ is then
\begin{mathletters}
\begin{eqnarray}
{\partial  W \over \partial \ln \Omega} =&& A P \, , \\
{\partial  P \over \partial \ln \Omega} =&&
{1 \over H^2 \Omega^2  } \widetilde D^2 A^{-1} W \, .
\label{ev3}
\end{eqnarray}
\end{mathletters}
In general, $(W,P)$ may be complex.

\subsection{Symmetry Conditions}

{}From the onset, it has been assumed that the scalar
matrix $S$ appearing in eq.(\ref{SRICCATI}) is symmetric. This
requirement will impose several conditions on the
fields appearing in the Riccati equation (\ref{riccs}).
Since
\begin{equation}
{\vec a}_1^T {\vec b}_2 = {\vec b}_1^T {\vec a}_2 \, ,
\end{equation}
one may show  that $A^{-1}\; C$ is indeed symmetric.
$P W^{-1}$ is symmetric provided that
\begin{equation}
W^T P - P^T W = 0 \, . \label{ic1}
\end{equation}
Using the first order evolution equation (\ref{ev3}), one verifies that
the derivative of the left-hand-side with respect to $\ln \Omega$
indeed vanishes. Hence, one typically imposes eq.(\ref{ic1}) as an initial
condition.

\subsection{Conjugate Conditions}

In addition, it is possible to arrange that
$(W,P)$ are canonically conjugate.
The conjugate conditions are
\begin{mathletters}
\begin{eqnarray}
W W^\dagger - W^* W^{T} =&& 0 \, ,  \label{identa} \\
P P^\dagger - P^* P^{T} =&& 0 \, ,  \label{identb} \\
W P^\dagger - W^* P^{T} =&& -i I \, , \label{identc}
\end{eqnarray}
\end{mathletters}
where $\dagger$ represents  the Hermitean conjugate,
$W^\dagger= (W^*)^{T}$, and $I$ is the 2-by-2 identity matrix,
$I_{ij} = \delta_{ij}$. If these conditions are met at one
time, then they are true for all times. For example,  one may thus verify
that the derivative of the left-hand-side with respect to $\ln \Omega$
vanishes (weakly).

Quantum considerations imply the conjugate conditions, which
legislate that $(W,P)$ are intrinsically complex since $i = \sqrt{-1}$
appears in eq.(\ref{identc}). The conjugate conditions
will be used to prepare the cosmological system in the ground
state at short wavelengths (see Sec. C).

The probability functional (\ref{prob}) for scalar perturbations
is determined to be
\begin{equation}
{\cal P} = {\rm exp} - {1 \over 2 } \int d^3x  f^{1/2} \;
[ \widetilde D^{-2} \widetilde R, e]
\; \left ( W W^\dagger \right )^{-1} \;
[ \widetilde D^{-2} \widetilde R, e]^T \, . \label{prob1}
\end{equation}
This functional is peaked about a flat and homogeneous
field configuration: $ \widetilde R=0 = \widetilde D^2 e$.
For small deviations about this configuration,
the 2-point correlation functions are found to be:
\begin{eqnarray}
< \left[ \widetilde D^{-2} \widetilde R(x) \right ]
\left[ \widetilde D^{-2} \widetilde R(y) \right ]   > =&&
f^{-1/2}(x) \; \left [ W W^\dagger(x) \right ]_{11}
\; \delta^3(x-y) \, , \\
 < \left[ \widetilde D^{-2} \widetilde R(x) \right ] e(y) > =&&
f^{-1/2}(x) \; \left [ W W^\dagger(x) \right ]_{12} \,  \; \delta^3(x-y) \, ,\\
  < e(x) \; e(y)  > =&& f^{-1/2}(x) \; \left [ W W^\dagger(x) \right ]_{22}
\; \delta^3(x-y) \, .
\end{eqnarray}
(Eq.(\ref{prob1}) follows from the Riccati equation (\ref{riccs}) and
the conjugate conditions (\ref{identa}-c):
\begin{eqnarray}
S - S^* =&& { 1 \over 2 \widetilde D^4} A^{-1}
\left [ { \partial W \over \partial \ln \Omega } W^{-1} -
{ \partial W^* \over \partial \ln \Omega } \left( W^{-1}\right ) \right ] \\
=&& { 1 \over 2 \widetilde D^4}
\left [ P W^\dagger \left(W^\dagger\right)^{-1} W^{-1} -
        P^* W^T \left( W^T \right)^{-1} \left( W^{-1} \right )^* \right ] \, ,
\quad etc. ;
\end{eqnarray}
in addition, $A^{-1}C$ was assumed to be real.)

\subsection{Short-Wavelength Behavior}

One of the main assumptions of the inflationary model
is that short wavelength fluctuations originated in the ground
state. Analytic expressions are given for two scalar fields in such a
state.

The Laplacian operator $\widetilde D^2$ for the conformal
3-metric $f_{ij}$ appears throughout this work. Much of the
subsequent operator analysis  is simplified by introducing
eigenfunctions $u(x, k)$ with eigenvalues $-k^2$:
\begin{equation}
\widetilde D^2 u(x,k) = -k^2 u(x,k) \, .
\end{equation}
In the basis of the eigenvectors,
the operator $W$ describing the ground state \cite{SBB89} evolves according to
\begin{equation}
W(\Omega,e,k) = { 1 \over \sqrt {2k}  \Omega } \;
e^{ i k  \tau} \; X^{-1} \, ,
\quad k /(H \Omega) >> 1 \, ,
\end{equation}
for wavelengths much shorter than the Hubble radius, $ k /(H \Omega) >> 1$.
Conformal time $\tau$ is given by
\begin{equation}
\tau= \int^\Omega \,  d {\overline \Omega} \, {\overline \Omega}^{-2} \;
{1 \over H( \overline \Omega, {e}) } \, ,
\end{equation}
and the components of the transformation matrix $X$ are
\begin{mathletters}
\begin{equation}
X_{11} = { 1 \over 4 } {\partial \phi_1 \over \partial \ln \Omega} \, , \quad
X_{12} =  {\partial \phi_1 \over \partial e} \, ,
\end{equation}
\begin{equation}
X_{21} = { 1 \over 4 } {\partial \phi_2 \over \partial \ln \Omega} \, , \quad
X_{22} =  {\partial \phi_2 \over \partial e} \, .
\end{equation}
\end{mathletters}
$X$ is actually the Jacobi matrix for the change of variables,
$( 4 \ln \Omega, e)  \rightarrow (\phi_1, \phi_2)$.
The inverse $X^{-1}$
\begin{mathletters}
\begin{equation}
(X^{-1})_{11} = 4 {\partial \ln \Omega \over \partial \phi_1 } \, , \quad
(X^{-1})_{12} = 4 {\partial \ln \Omega \over \partial \phi_2 } \, ,
\end{equation}
\begin{equation}
(X^{-1})_{21} = {\partial e  \over \partial \phi_1 } \, , \quad
(X^{-1})_{22} = {\partial e  \over \partial \phi_2} \, ,
\end{equation}
\end{mathletters}
is the Jacobi matrix for the inverse transformation,
$( \phi_1, \phi_2) \rightarrow ( 4 \ln \Omega, e)$.
Since
\begin{equation}
A = { 1 \over \Omega^3 H} \; X^{-1} (X^{-1})^T
\end{equation}
follows from eq.(\ref{mata}),
one may readily verify the symmetry condition (\ref{ic1}) and the
conjugate conditions (\ref{identa}-c).

\subsection{Matching Long and Short Wavelength Fluctuations}

The short-wavelength approximation breaks down when the comoving wavelength
equals the Hubble radius, $k /(H \Omega) \sim 1$,
at which time $W(\Omega, e, k)$ quickly approaches a constant value
$W^\infty(e, k) \equiv W(\infty, e, k)$ given approximately by
\begin{equation}
W^\infty(e, k) \sim { H \over \sqrt{2 k^3} }   \; X^{-1} \Big|_{k= H \Omega} \,
{}.
\end{equation}
$W^\infty(e, k)$ contains virtually all of the observable properties of the
cosmological model.

For a single scalar field, $\phi_1= \phi$, one obtains the well-known
expression for the variable zeta, $\zeta$:
\begin{equation}
\zeta(k) \equiv {3 \over 4} W^\infty(k) = { 3 H \over \sqrt{2 k^3} }
{\partial \ln \Omega \over \partial \phi } \; \Bigg|_{k= H \Omega}
= {  3 H^2 \over \sqrt{2 k^3} } \Bigg / \left ( { \dot \phi \over N } \right )
\; \Bigg|_{k= H \Omega} \, .
\end{equation}
The power spectrum for $\zeta$ is
\begin{equation}
{\cal P}_\zeta(k) \equiv { k^3 \over 2 \pi^2} |\zeta(k)|^2 =
{ 9 \over 4 \pi^2 }  H^4 \Bigg / \left ( { \dot \phi \over N} \right )^2
\Bigg|_{k= H \Omega} \, . \label{psz1}
\end{equation}
For two scalar fields, the power spectrum for $\zeta$ is
\begin{equation}
{\cal P}_\zeta(k) \equiv { 9 \over 16 } { k^3 \over 2 \pi^2}
\left[ W^\infty (W^\infty)^\dagger \right ]_{11}  \, .
\label{psz2}
\end{equation}
Examples for two scalar fields will be given in the next section.

For scalar perturbations, the resulting cosmic microwave background
temperature anisotropy is related to zeta through
\begin{equation}
\left ( { \Delta T(x) \over T }\right )_S = - \zeta(x)/ 15 \, ,
\quad |x| = 11,700 \rm{Mpc}
\end{equation}
where $|x| = 11,700 \; \rm{Mpc}$ is the comoving radius for the
surface of last scattering, assuming that the present Hubble
parameter is $H_0 = 50$ km / s/ Mpc.

\section{DOUBLE INFLATION POWER SPECTRUM}

As an example of the quadratic curvature approximation,
one computes fluctuations arising from inflation
that give rise to galaxy clustering as well as microwave
background anisotropies. Consider the potential for two
scalar fields,
\begin{equation}
V(\phi_a) = V_1(\phi_1) + V_2(\phi_2) \; ,
\end{equation}
which is the sum of two separable potentials,
$V_1(\phi_1)$ and  $V_2(\phi_2)$. This potential has been
used to describe the double inflation model, where
$\phi_1$ dominates the energy density of the first
epoch of inflation, and $\phi_2$ dominates the second epoch.

Although exact solutions of the Hubble function $H(\phi_a)$
exist \cite{S91}, for illustrative purposes, I will be content
to utilize the well-known slow-roll approximation,
\begin{equation}
H_{SR}(\phi_a) = \left [ { V_1(\phi_1) + V_2(\phi_2) \over 3 } \right ]^{1/2}\;
,
\end{equation}
where one neglects partial derivatives of the Hubble function
in the SHJE (\ref{SHJE}).
This is one of those cases were finding a solution depending on
two homogeneous parameters is very difficult, and I will instead
integrate eq.(\ref{che}). One may verify the solution:
\begin{mathletters}
\begin{eqnarray}
\ln \Omega =&& - \int^{\phi_1}_0 \; d \phi_1^\prime
{ V_1(\phi_1^\prime) \over { \partial V_1 \over \partial \phi_1^\prime} }
- \int^{\phi_2}_0 \; d \phi_2^\prime
{ V_2(\phi_2^\prime) \over { \partial V_2 \over \partial \phi_2^\prime} } \, ,
\\
e =&& \int^{\phi_1} { d \phi_1^\prime
\over { \partial V_1 \over \partial \phi_1^\prime} } -
\int^{\phi_2} { d \phi_2^\prime
\over { \partial V_2 \over \partial \phi_2^\prime} } \, .
\end{eqnarray}
\end{mathletters}
The inverse transformation matrix is
\begin{mathletters}
\begin{equation}
(X^{-1})_{11} = - 4 { V_1 \over { \partial V_1 \over \partial \phi_1} } \, ,
\quad
(X^{-1})_{12} = - 4 { V_2 \over { \partial V_2 \over \partial \phi_2} } \, ,
\end{equation}
\end{mathletters}
\begin{equation}
(X^{-1})_{21} =   {1 \over {\partial V_1 \over \partial \phi_1} } \, ,
\quad
(X^{-1})_{22} = - {1 \over {\partial V_2 \over \partial \phi_2} } \, .
\end{equation}
In the slow roll approximation, $k = H \Omega$ may be approximated
by $k \propto \Omega$.
The power spectrum, ${\cal P}_\zeta(k)$, eq.(\ref{psz2}), is then given by:
\begin{eqnarray}
{\cal P}_\zeta(k)=&& { 3 \over 4 \pi^2} \, (V_1+ V_2) \,
\left[
\left( V_1 \Big /  { \partial V_1 \over \partial \phi_1} \right )^2 +
\left( V_2 \Big /  { \partial V_2 \over \partial \phi_2} \right )^2
\right ] \, , \\
\ln \left[ k/ (10^{-4} {\rm Mpc}^{-1}) \right ]=&& 60
- \int^{\phi_1}_0 \; d \phi_1^\prime
{ V_1(\phi_1^\prime) \over { \partial V_1 \over \partial \phi_1^\prime} }
- \int^{\phi_2}_0 \; d \phi_2^\prime
{ V_2(\phi_2^\prime) \over { \partial V_2 \over \partial \phi_2^\prime} } \, ,
\\
e =&& \int^{\phi_1} { d \phi_1^\prime
\over { \partial V_1 \over \partial \phi_1^\prime} } -
\int^{\phi_2} { d \phi_2^\prime
\over { \partial V_2 \over \partial \phi_2^\prime} } \, .
\end{eqnarray}
It is assumed that the scale $k_0 = 10^{-4}$Mpc experienced
60 e-foldings of inflation after it crossed the Hubble radius
during inflation. For simplicity, $e$ is assumed to be a
homogeneous parameter. (Inhomogeneities in $e$ lead to
isocurvature perturbations which have been considered
by Sasaki and Yokoyama \cite{SY91}.)

These analytic results verify the numerical simulations of
ref.\cite{SBB89}. Heuristic arguments for these results
have been given previously in refs.\cite{Starobinsky85} and \cite{SB}.
The double inflation model is strongly constrained by cosmological
observations \cite{PETER94}.

The case of quadratic scalar field potentials,
\begin{equation}
V_1(\phi_1) = {1 \over 2} m_1^2  \phi_1^2 \,, \quad
V_2(\phi_2)=  {1 \over 2} m_2^2 \phi_2^2 \, , \label{V12}
\end{equation}
is particularly easy to implement.
In Fig.(1), power spectra for zeta are plotted for two models:
inflation with a single scalar
field, and  inflation with two scalar fields. Both models
have been normalized using COBE's 2-year data set \cite{BENNETT94}:
$\sigma_{sky}(10^0) \equiv [< (\Delta T)^2>]^{1/2} = 30.5 \pm 2.7 \mu K$.
Using the methods of refs. \cite{SS95}, \cite{S92},
a small correction has been made for gravitational waves.
For the single field model, the potential is
$V(\phi) =  m^2 \phi^2/ 2$, with $m= 5.45 \times 10^{-6}$
(recall that $m_P = \sqrt{ 8 \pi}$). For the double field model,
the potential is given by eq.(\ref{V12}), with $m_1 = 1.435 \times 10^{-5}$
and $m_2 = m_1/ 6$.

In Fig.(2), the observed galaxy data  are displayed using
the power spectrum for the density field,
$\delta = \delta \rho/ \rho$.  The  data were compiled
using several galaxy surveys \cite{PD94}, and it has been assumed
that infrared galaxies trace the mass distribution, {\it i.e.}, the
bias factor $b_{IRAS}=1$ is unity. For a single field, the theoretical
curve is unsatisfactory because it substantially overestimates the
power at short scales. For the double inflation model, the fit
is satisfactory provided the parameters have been carefully chosen.
Peter {\it et al} \cite{PETER94} have considered additional cosmological
tests. These models are the subject of a continuing investigation.

\section{CONCLUSIONS}

In elementary applications to mechanics, Hamilton-Jacobi theory
is useful because it enables one to compute the constants
of integration of a dynamical system.
This useful feature becomes even more attractive
when one considers Hamilton-Jacobi theory for general relativity.
Einstein's field equations contain two constraints which legislate that the
energy density and the momentum density vanish.  These constraints
are, in fact, constants of integration for the remaining evolution equations.
Hence by solving the Hamilton-Jacobi equation for general relativity,
one solves for the constraints as well as the other constants of integration
which include, for example, zeta, eq.(\ref{psz2}), for multiple
scalar fields.

Another attractive feature of HJ theory for general relativity
is that it yields a covariant formulation of gravity.
One may perform calculations without selecting an explicit choice
for the temporal and spatial coordinates. In fact, the generating
functional is manifestly invariant under transformations
of the spatial coordinates since it is constructed
by integrating  the Ricci tensor and spatial derivatives
of the various scalar fields over the 3-volume.
The invariance of the generating functional
under a change of time hypersurface is more subtle. Using a spatial
gradient expansion, one may express the generating functional
as a line integral in superspace. Each contour of integration in the
line integral corresponds to a  specific choice of time hypersurface,
but all choices yield the same answer. An integrability condition is met
because the generating functional is invariant under spatial
coordinate transformations.

For two scalar fields, one employs the method of characteristics
to transform the old variables $(\phi_1, \phi_2, f_{ij})$ to
new ones $(\Omega, e, f_{ij})$. The integration parameter
is arbitrary, but the choice of the conformal factor
$\Omega \equiv \Omega(\phi_a)$
simplifies the subsequent integration. The remaining new variables
$e$ and $f_{ij}$ are constants in the long-wavelength limit.
One may sum an infinite subset of terms in the spatial gradient
expansion by making an ansatz for the generating functional which
is quadratic in the fields $\widetilde D^2 e$, $\widetilde R$ and
$\widetilde R_{ij}$.

HJ methods have proved particularly useful in applications to
cosmology. As an illustration of the
theory, the fluctuation spectra for double inflation
was computed analytically confirming  previous work.
One may adjust the double inflation model to give
an adequate explanation of galaxy clustering as well as
large angle microwave background fluctuations.

\acknowledgments

The author thanks J.M. Stewart and D. Page for useful discussions. This
work was supported by the Natural Sciences and Engineering Research
Council of Canada, and the Canadian Institute for Theoretical Astrophysics.

\vfill\eject

\section{Figure Captions}

\noindent
{\bf Fig.(1)}: Power spectra for zeta are shown
for two models: (1) standard single field inflation and
(2) double inflation. Both utilize quadratic scalar field
potentials. The spectra are normalized using large
angle microwave background measurements. A small
correction has been made for gravitation radiation.
Double inflation gives less power at short scales.

\noindent
{\bf Fig.(2)}: The power spectrum for the observed
density field $\delta = \delta \rho / \rho$ is shown.
Standard single field inflation
gives a poor fit of the data at shorter distances,
$k < 10^{-1.4}$ Mpc$^{-1}$. By adjusting several
free parameters, double inflation removes this problem.
\vfill\eject

%
%
\begin{figure}
\setlength{\unitlength}{0.240900pt}
\ifx\plotpoint\undefined\newsavebox{\plotpoint}\fi
\sbox{\plotpoint}{\rule[-0.175pt]{0.350pt}{0.350pt}}%

\end{figure}
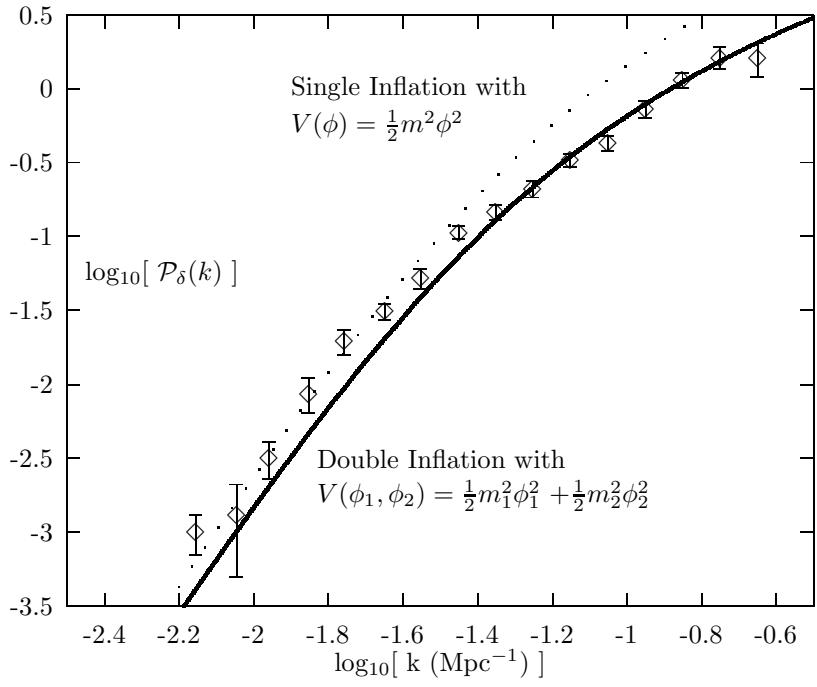
\end{document}